%% file: wan.tex
\documentclass{article}
\usepackage{spconf,amsmath,amssymb,graphicx,booktabs,multirow,siunitx,enumitem,placeins,float,balance}
\usepackage[hidelinks]{hyperref}

\makeatletter
\renewcommand{\normalsize}{%
  \@setfontsize\normalsize{9pt}{9.1pt}%
  \abovedisplayskip=4pt plus 1pt minus 1pt
  \belowdisplayskip=4pt plus 1pt minus 1pt
  \abovedisplayshortskip=2pt plus 1pt
  \belowdisplayshortskip=3pt plus 1pt minus 1pt
}
\makeatother
\makeatletter
\newcommand{\bstctlcite}[1]{%
  \@bsphack
  \@for\@citeb:=#1\do{%
    \if@filesw
      \immediate\write\@auxout{\string\citation{\@citeb}}%
    \fi
    \@ifundefined{b@\@citeb}{%
      \expandafter\gdef\csname b@\@citeb\endcsname{}%
    }{}%
  }%
  \@esphack
}
\makeatother
\newcommand{\laalcu}{\ensuremath{\mathrm{LAAL}_{\mathrm{CU}}}}
\setlist[itemize]{topsep=3pt,itemsep=1pt,parsep=0pt,partopsep=0pt}
\title{Persistent Delivery Optimization for Streaming Speech-to-Text Translation with Revisions}
\name{Zixiang Wan$^{1}$, Delin Chen$^{1}$, Wei Shi$^{2}$, Haihua Xu$^{2}$, Youxi Xie$^{2}$, Yuexian Zou$^{1}$\sthanks{Corresponding author.}}
\address{$^{1}$Peking University, Shenzhen, China\\
$^{2}$Timekettle AI Lab, Shenzhen, China\\
\texttt{zxwan25@stu.pku.edu.cn}, \texttt{zouyx@pku.edu.cn}}

\begin{document}
\normalsize

\maketitle
\bstctlcite{IEEEexample:BSTcontrol}

\begin{abstract}
Revision-capable streaming speech-to-text translation (S2TT) can correct
earlier drafts, but process rewards based on visible text may credit content
later withdrawn. Persistent Delivery Optimization (PDO) assigns intermediate
reward only to content that survives revisions while scoring final quality
separately. With 7.49 h of task-specific FLEURS adaptation, PDO achieves the
best BLEU on four of five directions and higher COMET than every external
streaming baseline in all five directions.
Relative to its History-SFT initialization, PDO reduces
mean/P90 finalization-aware latency by 10.8\%/11.3\% and normalized
erasure by 15.8\%, while emitting at the first permitted 2-s update
and improving macro BLEU. Zero-shot evaluation on Europarl-ST and CoVoST 2
confirms that these gains are not confined to the FLEURS training domain.\footnotemark[1]
\end{abstract}


\footnotetext[1]{Training code and model weights:
\href{https://github.com/ggiggit/PDO_S2TT}{github.com/ggiggit/PDO\_S2TT}}

\section{Introduction}
\label{sec:intro}

End-to-end streaming speech-to-text translation (S2TT) directly generates
target-language text from incoming source speech and must balance translation
quality against latency. Emitting early relies on incomplete source context and
can increase translation uncertainty, whereas waiting provides more source
evidence at the cost of higher latency. In conventional monotonic streaming
systems, emitted target tokens are irrevocable, so early translation errors
cannot be corrected. Allowing revisions relaxes this constraint by letting the
system update previously visible drafts as more source speech becomes
available~\cite{arivazhagan2020retranslation,arivazhagan2020strategies}.

Revision separates first appearance from finalization. In monotonic streaming
translation, emitted content is irrevocable, so the current visible draft
reflects content that has already been committed. With revision, previously
displayed text may later be changed or removed. A process reward computed on the
current visible draft can therefore reward temporary content that does not
persist in the final translation.

To avoid assigning early-output credit to temporary content, we reward only
the content that appears early and remains unchanged through subsequent
revisions. For each intermediate visible draft, we identify the longest prefix
preserved in all later drafts and accumulate a reference-coverage score
for this persistent content over time,
while evaluating final translation quality separately. Content that becomes
stable earlier therefore receives greater reward, whereas text that is later
revised or removed does not retain the same benefit from its early appearance.

Our contributions are summarized as follows:
\begin{itemize}
    \item We identify a mismatch between visibility and delivery in streaming
    translation with revisions. Process rewards based on the current visible draft may
    assign early-output credit to content that is later withdrawn.

    \item We propose Persistent Delivery Optimization (PDO), a trajectory-level
    objective that accumulates the reference coverage of content retained through
    subsequent revisions and combines it with final translation quality.
    Full-trajectory returns allow later revisions to affect the credit assigned
    to earlier outputs.

    \item We instantiate PDO in a history-conditioned end-to-end streaming
    S2TT system without an intermediate source transcript. Controlled reward
    comparisons isolate the benefit of persistent-content credit. Zero-shot
    evaluation on two out-of-domain corpora shows that the gains over
    History-SFT are not confined to FLEURS or obtained by delaying the first
    response.
\end{itemize}

\section{Related Work}
\label{sec:related}

\subsection{End-to-End Streaming Speech-to-Text Translation}
\label{ssec:related_simulst}

End-to-end streaming S2TT generates target-language text directly from speech.
Early monotonic systems include SimulSpeech~\cite{ren2020simulspeech}, which
combines online segmentation with wait-k decoding, and
SimulMT-to-SimulST~\cite{ma2020simulmt}, which applies SimulMT policies through
pre-decision mechanisms. Adaptive approaches such as EDAtt
~\cite{papi2023attention} and AlignAtt~\cite{papi2023alignatt} use
audio--translation attention to decide when to emit, while
SimulSeamless~\cite{papi2024simulseamless} scales such policies to large
pretrained speech translation models.

Policy-learning methods further optimize quality and latency. Hibiki-Zero
~\cite{labiausse2026hibikizero} uses intermediate translation-quality rewards
without word-level alignment, while HPO~\cite{ouyang2026hpo} jointly optimizes
both objectives on unbounded speech. PDO instead targets a revision-specific
failure: visible-text credit may remain after that text is withdrawn.

\subsection{Revision and Information Delivery}
\label{ssec:related_revision}

Revision-based translation updates visible drafts as source context arrives.
Prior work stabilizes re-translation through masking and biased decoding
~\cite{arivazhagan2020strategies,arivazhagan2020retranslation}, self-training
for reduced flicker~\cite{sen2023selftraining}, and constraints on the revision
range~\cite{arivazhagan2020strategies}. Finalization and erasure metrics measure
when output stabilizes~\cite{papi2022laal} and how much displayed content is
later withdrawn~\cite{arivazhagan2020strategies}.

Stability, however, is not the objective by itself. Delaying emission can make
a system appear stable, while a necessary revision may reduce stability yet
improve the information shown to the user. PDO therefore studies when
translated information becomes both available and persistent under subsequent
revisions, rather than simply minimizing changes.
\section{Method}
\label{sec:method}

We adapt Qwen3-ASR-1.7B~\cite{shi2026qwen3asr} into a history-conditioned streaming S2TT
policy, then optimize persistent delivery with PDO.

\subsection{Streaming S2TT Policy}
\label{ssec:streaming_policy}

We initialize from Qwen3-ASR-1.7B and freeze its pretrained backbone.
Let $x$, $x^{\mathrm{txt}}$, and $y^{(\ell)}$ denote source speech, its
training transcript, and the reference in target language $\ell$. We train
target-task LoRA~\cite{hu2022lora} $\phi$ followed by history module $\eta$, without an additional ASR
stage or intermediate source transcripts at inference.

Supervised adaptation proceeds through text translation, full-audio S2TT,
source-text conditioning reduction, and streaming SFT. To reduce reliance on
training transcripts, the source-text conditioning reduction stage mixes full, corrupted,
prefix, and empty source-text conditions with weights
$(0.4,0.3,0.2,0.1)$; the empty condition provides
direct speech-to-target supervision.

On the update grid $0<u_1<\cdots<u_T=U$, streaming SFT combines
 a complete-translation anchor with cumulative drafts from Hibiki-style
contextual alignment~\cite{labiausse2026hibikizero} using Qwen3-ASR forced
word times~\cite{shi2026qwen3asr}. Token-normalized
complete and streaming supervision receive equal weight; draft events
are averaged within each target direction, and directions sharing a
recording are weighted equally. Alignment-derived drafts are
re-tokenized with the model tokenizer. Source--target alignments and
source-word timing are used only to construct streaming supervision,
not at inference. Empty early drafts supervise only EOS.

\noindent\textbf{History-conditioned SFT.}
We freeze the trained streaming SFT policy and perform a closed-loop
rollout over the training set to cache model-generated histories. Let
$d_0=\varnothing$. At update $t$, its previous visible draft $d_{t-1}$, rather than a
reference prefix, supplies the history. These cached histories provide
the supervised states for History-SFT: the backbone and target-task
LoRA remain frozen, and only $\eta$ is trained. History is injected
into the upper decoder layers through zero-initialized gated
cross-attention. History-SFT applies a first-order projection that prevents the
history-module update from increasing the frozen parent policy's continuation loss.
The resulting policy $\pi_{\mathrm{hist}}$ operates on
$s_t=(x_{\le u_t},d_{t-1},\ell)$ and generates token sequence
$z_t\sim\pi_{\mathrm{hist}}(\cdot\mid s_t)$. Each update decodes $z_t$
into a complete target draft $d_t$ that replaces the previous visible
draft. PDO initializes from $\pi_{\mathrm{hist}}$ and updates
$\theta=(\phi,\eta)$.

\subsection{Persistent Delivery Optimization}
\label{ssec:pdo}

\noindent\textbf{Persistent-delivery utility.}
For target language $\ell$, write $y=y^{(\ell)}$ and let
$\bar{d}_t=\nu_\ell(d_t)$ and $\bar{y}=\nu_\ell(y)$ denote case-folded text-unit
sequences after discarding surrounding punctuation but retaining intra-word
apostrophes and hyphens. Units are Han/kana characters plus contiguous non-CJK
alphanumeric spans for Chinese/Japanese and Unicode words otherwise. For
$\tau=(d_1,\ldots,d_T)$, the content visible at update $t$ that
survives all subsequent revisions is
\begin{equation}
p_t=\operatorname{LCP}(\bar{d}_t,\bar{d}_{t+1},\ldots,\bar{d}_T).
\label{eq:persistent}
\end{equation}
By construction, $p_t\preceq p_{t+1}$, where $\preceq$ denotes the
prefix relation. With $\Delta_t=u_{t+1}-u_t$, we define
\begin{equation}
J_{\mathrm{PD}}(\tau)=
\sum_{t=1}^{T-1}\Delta_t q_p(p_t,\bar{y})
+Hq_s(\bar{d}_T,\bar{y}),
\label{eq:jpd}
\end{equation}
where
$q_p(p_t,\bar{y})=|\operatorname{LCS}(p_t,\bar{y})|/|\bar{y}|$
is persistent-prefix lexical coverage (ROUGE-L recall~\cite{lin2004rouge}),
and $q_s$ is normalized effective-order sentence BLEU~\cite{post2018sacrebleu};
$H>0$ weights terminal quality. Earlier-persistent content accrues more
source-time credit, whereas withdrawn text receives no such early-display
benefit; monotonic output has $p_t=\bar{d}_t$.

\noindent\textbf{Trajectory credit.}
To let later revisions affect the credit assigned to earlier drafts,
we define
\begin{equation}
\begin{aligned}
\Phi_t={}&\sum_{k=1}^{t-1}\Delta_k
q_p\!\left(\operatorname{LCP}(\bar{d}_k,\ldots,\bar{d}_t),\bar{y}\right)\\
&{}+Hq_s(\bar{d}_t,\bar{y}),\\
G_t={}&\Phi_T-\Phi_{t-1},
\end{aligned}
\label{eq:return}
\end{equation}
with $\Phi_0=0$ and $\Phi_T=J_{\mathrm{PD}}(\tau)$.
Equivalently, $G_t=\sum_{t'=t}^{T}r_{t'}$ for
$r_t=\Phi_t-\Phi_{t-1}$. Returns are computed after a complete
rollout, while inference remains causal.

\noindent\textbf{Policy update.}
PDO proceeds in rollout rounds using PPO-style clipped updates~\cite{schulman2017ppo}
with group-relative standardized returns~\cite{shao2024deepseekmath}. Each round freezes the current policy
as the proximal snapshot $\pi_{\mathrm{prox}}$. A temperature-adjusted
behavior distribution $\pi_b$, induced by the same round-start policy,
samples $K$ fresh closed-loop trajectories per utterance: every generated
draft supplies the history for the next speech update. After all
trajectories finish, we compute returns and standardize them within
each utterance and update as
$w_t^{(i)}=(G_t^{(i)}-\mu_t)/(\sigma_t+\varepsilon_{\mathrm{num}})$.
Here, $\mu_t$ and $\sigma_t$ are the group mean and standard deviation,
and $\varepsilon_{\mathrm{num}}$ is a numerical stabilizer;
near-zero-variance groups receive zero weight.

The trajectories and $\pi_{\mathrm{prox}}$ remain fixed during the
round's optimizer updates to $\pi_\theta$. The updated policy starts
the next round, refreshing both the snapshot and trajectories.
For token $z_{t,j}^{(i)}$ under generation context $h_{t,j}^{(i)}$, define
$\rho_{t,j}^{(i)}=\pi_\theta/\pi_{\mathrm{prox}}$ and
$c_{t,j}^{(i)}=\allowbreak\min(c_{\max},\allowbreak\pi_{\mathrm{prox}}/\pi_b)$,
with all probabilities evaluated at $(z_{t,j}^{(i)}\mid h_{t,j}^{(i)})$.
The ratio $\rho$ constrains updates relative to the round-start snapshot,
while $c$ gives a truncated correction for temperature-adjusted sampling.
For trajectory minibatch $\mathcal{B}$, we optimize
\begin{equation}
\begin{aligned}
\mathcal{L}_{\mathrm{PDO}}={}&-\frac{1}{|\mathcal{B}|}
\sum_{i,t,j}c_{t,j}^{(i)}\\
&{}\times\min\!\Bigl[\rho_{t,j}^{(i)}w_t^{(i)},\\
&\quad\operatorname{clip}\bigl(\rho_{t,j}^{(i)},
1-\varepsilon_{\mathrm{clip}},1+\varepsilon_{\mathrm{clip}}\bigr)
w_t^{(i)}\Bigr].
\end{aligned}
\label{eq:pdo}
\end{equation}
All tokens in a draft share its event weight $w_t^{(i)}$, while
the sampling correction and policy ratio are token-specific.
The loss is normalized by the number of trajectories rather than the
number of generated tokens. PDO updates $\phi$ and $\eta$ while the
pretrained backbone remains frozen.

\section{Experiments}
\label{sec:experiments}

\subsection{Experimental Setup}
\label{ssec:experimental_setup}

\input{tables/fleurs_results}
\input{tables/streaming_system_profiles}

\noindent\textbf{Data.}
We train only on FLEURS~\cite{conneau2022fleurs} En$\rightarrow$Zh/De/Es/Ja/Fr.
TRAIN, DEV, and TEST contain 2,602/\allowbreak394/\allowbreak647 shared English
recordings (7.49/\allowbreak1.05/\allowbreak1.77 h); targets come from the n-way-parallel
references for the same sentence IDs. Streaming stages use the 2,600 TRAIN
recordings with valid timing supervision; models are selected on DEV and
evaluated on TEST. For zero-shot cross-domain evaluation, the same checkpoints
are tested on the official Europarl-ST
v1.1~\cite{iranzo2020europarlst} TEST sets for En$\rightarrow$De/Es/Fr with
original segmentation, and CoVoST 2~\cite{wang2020covost2} for En$\rightarrow$Zh/De/Ja.
History-SFT and PDO use the same 2-s update grid, with no target-domain
adaptation.

\noindent\textbf{Baselines.}
We compare with SeamlessStreaming~\cite{seamless2023}, EASiST~\cite{fu2026easist},
Hikari-medium~\cite{koshkin2026hikari}, and the text path of
SimulS2ST-Omni~\cite{he2026simuls2stomni} as direct streaming systems, and
with SimulStreaming~\cite{machacek2025simulstreaming} and
AlignAtt4LLM~\cite{fuxa2026alignatt4llm} as transcription-mediated controls.
We additionally use Qwen3-ASR-1.7B~\cite{shi2026qwen3asr}$\rightarrow$
Hy-MT2-1.8B~\cite{zheng2026hymt2} as a matched-backbone full-context cascade.
We evaluate EASiST Stage-3, SimulStreaming with Whisper
large-v3~\cite{radford2022whisper} + EuroLLM-1.7B~\cite{martins2025eurollm},
and the canonical AlignAtt4LLM-Gemma configuration.

\noindent\textbf{Controlled RL baselines.}
All variants start from History-SFT and share the state/action space, rollout
and update budget, sampling configuration, updater, and trainable parameters.
Current-draft RL uses current-draft rather than persistent-prefix coverage
throughout return construction; Hibiki-Zero-style~\cite{labiausse2026hibikizero} uses
intermediate-plus-final BLEU ($\alpha=.5$); HPO-style~\cite{ouyang2026hpo} uses quality-gated
final-BLEU/\laalcu{} with threshold .33 and latency weight .5.

\input{tables/reward_design_ablation}
\input{tables/cross_domain_results}

\noindent\textbf{Metrics.}
We report SacreBLEU 2.6.0~\cite{post2018sacrebleu} (\texttt{zh} for Chinese,
\texttt{ja-mecab} with MeCab 0.996 and ipadic for Japanese, and \texttt{13a}
otherwise), chrF++~\cite{popovic2017chrf} (word order 2), and
WMT22-COMET-DA~\cite{rei2022comet22}, with COMET scaled by 100; macros are
unweighted direction averages. First-token latency (FTL) is source audio
consumed until the first non-empty output; final-response delay (FRD) adds
inference and queueing time. Our finalization-aware, computation-unaware
length-adaptive average lagging (\laalcu{})~\cite{papi2022laal} timestamps each
final unit at the earliest source-consumption time after which its final prefix
never changes. Unless stated
otherwise, \laalcu{} is in seconds, written as mean / P90. For \laalcu{}
and revision metrics, target units are characters for Chinese/Japanese
and words otherwise.
Normalized erasure is erased target units per final-output unit~\cite{arivazhagan2020strategies}; age-weighted
erasure weights each erased unit by its display age. Real-time factor (RTF) is
inference time divided by source duration.

\noindent\textbf{Training.}
We freeze Qwen3-ASR-1.7B throughout adaptation. Target-task LoRA uses rank
16 ($\alpha=32$, dropout .05) on decoder $q/k/v/o$ projections; the history
module applies 256-dimensional four-head cross-attention to the top four
decoder layers with at most 128 history tokens. All streaming stages use a
2-s grid and proceed through text translation, full-audio S2TT,
source-text conditioning reduction, streaming SFT, and History-SFT.
History-SFT trains only $\eta$ from cached frozen-parent histories. PDO updates
$\phi$ and $\eta$ (15.9M parameters) with $K=4$, temperature .2, $H=2$ s,
$\varepsilon_{\mathrm{clip}}=.2$, and $c_{\max}=2$.

\subsection{In-Domain Results}
\label{ssec:main_results}

Table~\ref{tab:fleurs} establishes in-domain competitiveness. Among external
streaming systems, PDO has the best BLEU in four of five directions and higher
COMET in all five; on En$\rightarrow$De/Es, it also exceeds direct EASiST in
BLEU, COMET, and chrF++ while lowering FTL and FRD.

The matched History-SFT comparison isolates the effect of PDO: macro BLEU
increases while mean/P90 \laalcu{} and normalized erasure decrease. Because
both systems emit at the first permitted 2-s update, the finalization-aware
latency gain reflects earlier stabilization rather than delayed first output.

Table~\ref{tab:profile} contextualizes scale: PDO adapts 15.9M parameters on
7.49 h of task-specific S2TT. It also exceeds the matched-backbone full-context
cascade on all three reported quality metrics, showing that direct streaming
operation need not sacrifice final translation quality or introduce an
ASR-to-MT error-propagation interface.

\subsection{Ablation Study}
\label{ssec:ablation}

Table~\ref{tab:reward_ablation} tests persistent-content credit against matched
current-draft, intermediate-plus-final-BLEU, and quality-gated
BLEU/\laalcu{} rewards. Current-draft RL retains
quality but sharply increases FRD and erasure, showing that visible-content
reward can favor temporary output. HPO-style is the strongest alternative,
yet PDO has lower mean/P90 \laalcu{} and erasure with comparable FRD. This
controlled comparison attributes the delivery gains to persistent credit.

\subsection{Persistent Delivery Analysis}
\label{ssec:persistent_analysis}

\begin{figure}[H]
\centering
\includegraphics[width=\linewidth,trim=6bp 6bp 6bp 6bp,clip]{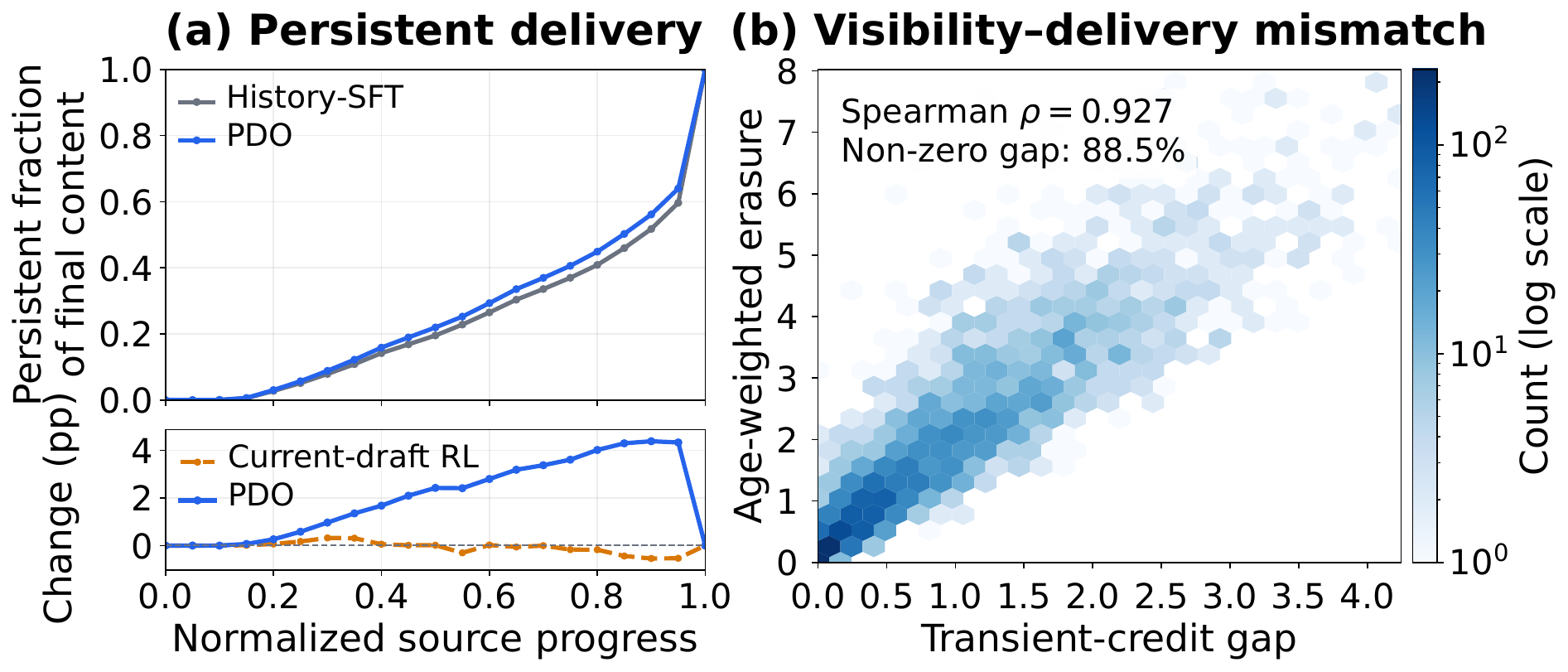}
\caption{FLEURS TEST persistent-delivery analysis. (a) Final-content fraction
and change from History-SFT (ending at zero by construction). (b)
Transient-credit gap versus age-weighted erasure ($\rho=.927$).}
\label{fig:persistent}
\end{figure}

Figure~\ref{fig:persistent} tests the proposed visibility--delivery mechanism
on FLEURS TEST. For normalized source progress
$s$, set $t(s)=\max(\{t:u_t/U\leq s\}\cup\{0\})$ and $p_0=\varnothing$.
We plot the persistent fraction $f(s)=|p_{t(s)}|/|\bar d_T|$, averaged over
3,235 direction-records on $s\in\{0,.05,\ldots,1\}$. We define the
transient-credit gap as excess current-draft coverage credit:
\[
\begin{aligned}
g(\tau)&=\sum_{t=1}^{T-1}\Delta_t
  \left[q(\bar d_t,\bar y)-q(p_t,\bar y)\right],
\end{aligned}
\]
where $q(a,\bar y)=|\operatorname{LCS}(a,\bar y)|/|\bar y|$; terminal rewards
cancel in $g$. Figure~\ref{fig:persistent}(a) shows that PDO makes more final
content stably available earlier, whereas Current-draft RL remains near
History-SFT. Figure~\ref{fig:persistent}(b) shows non-zero transient credit in
88.5\% of 3,235 PDO trajectories and strong correlation with age-weighted
erasure ($\rho=.927$). Together, these patterns show that current-draft rewards
often credit temporary content and that this mismatch tracks revision
instability.

\subsection{Zero-Shot Cross-Domain Evaluation}
\label{ssec:cross_domain}

Because Table~\ref{tab:fleurs} evaluates FLEURS-adapted models on FLEURS,
Table~\ref{tab:crossdomain} tests domain specificity using the same checkpoints
without target-domain adaptation. History-SFT isolates PDO's effect;
SeamlessStreaming is a multilingual direct external reference covering all six
directions. PDO improves BLEU over History-SFT in every direction and reduces
finalization-aware latency and erasure with unchanged first-update timing. Its
higher macro BLEU and COMET than SeamlessStreaming on both corpora confirm that
PDO's advantages are not artifacts of in-domain FLEURS evaluation but
generalize zero-shot to unseen domains.

\section{Conclusion}
\label{sec:conclusion}

Revision-capable streaming S2TT may reward early text later withdrawn. PDO
corrects this mismatch by crediting reference-matching content only if it
survives later revisions, while scoring final quality separately. Controlled
ablations and trajectory analysis show that persistence-aware credit---not
generic RL---enables earlier stable delivery with less revision instability.
Across five FLEURS directions, PDO improves macro BLEU while reducing
finalization-aware latency and erasure relative to History-SFT, without
delaying first emission. Zero-shot gains on Europarl-ST and CoVoST 2 show that
the benefit generalizes beyond the training domain. Together, these results
support optimizing persistent information rather than transient visibility in
revision-capable streaming translation.

\newpage
\balance
\bibliographystyle{IEEEtran}
\bibliography{references}

\end{document}

%% file: tables/fleurs_results.tex
\setcounter{table}{1}%
\begin{table*}[!t]
\centering
\caption{In-domain FLEURS TEST results. Latency columns are first-token latency (FTL),
final-response delay (FRD), finalization-aware computation-unaware
length-adaptive average lagging (\laalcu{}; mean/P90), and real-time factor
(RTF). Bold/underline mark best/second-best per language or macro block; ties
share formatting. Each \laalcu{} pair shares one rank; RTF and the full-context
reference are unranked. History-SFT/PDO are three-seed means.}
\label{tab:fleurs}
\setlength{\tabcolsep}{0pt}
\renewcommand{\arraystretch}{.86}
\begin{tabular*}{\textwidth}{@{\extracolsep{\fill}}clrrrrrcr@{}}
\toprule
& & \multicolumn{3}{c}{\textbf{Translation Quality}}
& \multicolumn{3}{c}{\textbf{Latency (s)}}
& \\
\cmidrule(lr){3-5}\cmidrule(lr){6-8}
\textbf{Dir.} & \textbf{System}
& BLEU$\uparrow$ & COMET$\uparrow$ & chrF++$\uparrow$
& FTL$\downarrow$ & FRD$\downarrow$ & \laalcu$\downarrow$
& RTF \\
\midrule
\multirow{6}{*}{Zh} & SeamlessStreaming & 26.85 & 76.79 & 18.08 & 2.48 & 2.60 & 2.36 / 3.42 & .14 \\
 & SimulS2ST-Omni & 34.13 & 80.83 & 25.31 & 2.25 & 2.97 & \underline{1.70 / 2.74} & .33 \\
 & SimulStreaming & 12.60 & 72.83 & 14.59 & \textbf{1.89} & 2.45 & 2.38 / 4.04 & .57 \\
 & AlignAtt4LLM & 27.14 & 75.69 & 22.32 & 3.01 & 3.54 & \textbf{1.24 / 2.09} & .38 \\
 & History-SFT & \underline{37.60} & \textbf{86.72} & \textbf{26.98} & \underline{2.00} & \underline{2.36} & 4.03 / 7.47 & .43 \\
 & \textbf{PDO} & \textbf{38.16} & \underline{86.31} & \underline{26.92} & \underline{2.00} & \textbf{2.33} & 3.53 / 6.61 & .39 \\
\midrule
\addlinespace[1pt]
\multirow{7}{*}{De} & SeamlessStreaming & \underline{28.30} & 81.85 & 55.97 & 2.53 & 2.66 & 1.93 / 2.80 & .15 \\
 & Hikari-medium & 19.49 & 68.02 & 48.83 & 2.10 & \underline{2.10} & \textbf{0.90 / 1.76} & 1.43 \\
 & SimulStreaming & 16.00 & 75.68 & 49.24 & \textbf{1.42} & \textbf{1.64} & 2.01 / 3.03 & .30 \\
 & EASiST & 24.21 & 76.54 & 53.20 & 5.70 & 6.06 & 4.35 / 9.52 & .16 \\
 & AlignAtt4LLM & 25.34 & 74.18 & \underline{56.67} & 3.01 & 3.57 & \underline{1.50 / 2.27} & .41 \\
 & History-SFT & 28.10 & \textbf{84.48} & 55.85 & \underline{2.00} & 2.47 & 3.26 / 6.00 & .62 \\
 & \textbf{PDO} & \textbf{29.57} & \underline{84.22} & \textbf{57.03} & \underline{2.00} & 2.42 & 3.08 / 5.65 & .57 \\
\midrule
\addlinespace[1pt]
\multirow{5}{*}{Es} & SeamlessStreaming & 22.67 & 81.60 & 50.10 & 2.80 & 2.98 & \textbf{1.78 / 2.61} & .14 \\
 & SimulStreaming & 17.80 & 76.83 & 46.45 & \textbf{1.56} & \textbf{1.75} & \underline{2.06 / 3.16} & .27 \\
 & EASiST & 19.86 & 75.35 & 46.92 & 4.44 & 4.71 & 3.10 / 5.63 & .15 \\
 & History-SFT & \underline{22.88} & \underline{83.56} & \underline{50.30} & \underline{2.00} & 2.46 & 3.13 / 6.06 & .58 \\
 & \textbf{PDO} & \textbf{23.35} & \textbf{83.57} & \textbf{50.88} & \underline{2.00} & \underline{2.44} & 2.51 / 4.44 & .57 \\
\midrule
\addlinespace[1pt]
\multirow{5}{*}{Ja} & SeamlessStreaming & 16.01 & 80.56 & 20.13 & 3.01 & 3.19 & 3.55 / 5.34 & .15 \\
 & Hikari-medium & 20.26 & 76.96 & 21.24 & 2.43 & \underline{2.43} & \underline{2.12 / 3.67} & 1.41 \\
 & SimulStreaming & 9.55 & 73.47 & 16.67 & \textbf{1.24} & \textbf{1.97} & \textbf{1.62 / 3.10} & .65 \\
 & History-SFT & \underline{28.30} & \textbf{88.55} & \underline{26.24} & \underline{2.00} & 2.49 & 3.98 / 7.02 & .58 \\
 & \textbf{PDO} & \textbf{28.66} & \underline{88.50} & \textbf{26.47} & \underline{2.00} & 2.47 & 3.67 / 6.23 & .57 \\
\midrule
\addlinespace[1pt]
\multirow{4}{*}{Fr} & SeamlessStreaming & \textbf{38.95} & 82.77 & \underline{61.56} & 2.54 & 2.69 & \textbf{1.69 / 2.52} & .15 \\
 & SimulStreaming & 26.26 & 76.77 & 55.86 & \textbf{1.48} & \textbf{1.66} & \underline{1.92 / 2.89} & .28 \\
 & History-SFT & 36.95 & \textbf{84.72} & 60.97 & \underline{2.00} & 2.46 & 2.70 / 5.18 & .61 \\
 & \textbf{PDO} & \underline{38.16} & \underline{84.26} & \textbf{62.06} & \underline{2.00} & \underline{2.45} & 2.46 / 4.93 & .59 \\
\midrule
\addlinespace[1pt]
\multicolumn{9}{l}{\textit{Five-direction macro}} \\
 & SeamlessStreaming & 26.56 & 80.71 & 41.17 & 2.67 & 2.82 & \underline{2.26 / 3.34} & .15 \\
 & SimulStreaming & 16.44 & 75.12 & 36.56 & \textbf{1.52} & \textbf{1.90} & \textbf{2.00 / 3.24} & .41 \\
 & History-SFT & \underline{30.77} & \textbf{85.61} & \underline{44.07} & \underline{2.00} & 2.45 & 3.42 / 6.38 & .56 \\
 & \textbf{PDO} & \textbf{31.58} & \underline{85.37} & \textbf{44.67} & \underline{2.00} & \underline{2.42} & 3.05 / 5.66 & .54 \\
\midrule
\addlinespace[1pt]
\multicolumn{9}{l}{\textit{Full-context reference}} \\
& Qwen3-ASR$\rightarrow$Hy-MT2 & 30.79 & 82.90 & 44.64
& \multicolumn{3}{c}{\textit{full context}} & .13 \\
\bottomrule
\end{tabular*}
\vspace{0pt}
\parbox{\textwidth}{\raggedright\normalsize
SimulStreaming En$\rightarrow$Zh: quality coverage 94.6\%; FTL/FRD and
\laalcu{} use 96.3\%/94.3\% valid subsets.}
\end{table*}

%% file: tables/streaming_system_profiles.tex
\setcounter{table}{0}%
\begin{table}[H]
\centering
\nocite{seamless2023,machacek2025simulstreaming,fu2026easist,koshkin2026hikari,fuxa2026alignatt4llm,he2026simuls2stomni}
\caption{Streaming-system profiles and training resources~\cite{seamless2023}--\cite{he2026simuls2stomni}.}
\label{tab:profile}
\normalsize
\setlength{\tabcolsep}{0pt}
\renewcommand{\arraystretch}{.92}
\begin{tabular*}{\linewidth}{@{\extracolsep{\fill}}p{84pt}@{\hspace{-10pt}}lcS[table-format=2.2,table-number-alignment=center]p{65pt}@{}}
\toprule
\textbf{System} & \textbf{Yr.} & \textbf{Mode}
& \multicolumn{1}{c}{\textbf{Params}} & \textbf{Resources} \\
\midrule
SeamlessStreaming & 2023 & Direct & 2.50 & 351K+145K h \\
SimulStreaming & 2025 & Src-text & 3.20 & pretr. ASR+MT \\
EASiST & 2026 & Direct & 8.36 & MuST-C SimulST \\
Hikari-medium & 2026 & Direct & 0.80 & 62.3K h \\
AlignAtt4LLM & 2026 & Src-text & 11.26 & ASR+align.+MT \\
SimulS2ST-Omni & 2026 & Direct & 4.04 & 24.4K h+aux. \\
\midrule
\textbf{PDO} & --- & Direct & 2.35
& \textbf{7.49 h adapt.} \\
\bottomrule
\end{tabular*}
\vspace{1pt}
\parbox{\linewidth}{\raggedright\footnotesize
\textbf{Mode:} Direct = no source transcript at inference; Src-text = with one.
Params are in B; PDO updates 15.9M. \textbf{Hours:} Seamless = S2TT/ASR+S2ST;
Hikari = S2TT+ASR; Omni = S2TT+aux.; PDO = S2TT adaptation. Aux. includes
ASR/MT/TTS/S2ST.}
\end{table}
\setcounter{table}{2}%

%% file: tables/reward_design_ablation.tex
\begin{table}[!t]
\centering
\caption{Controlled FLEURS TEST reward ablation (five-direction macro). RL
variants share History-SFT initialization and update framework; all entries are
three-seed means.}
\label{tab:reward_ablation}
\setlength{\tabcolsep}{0pt}
\begin{tabular*}{\linewidth}{@{\extracolsep{\fill}}lrrrrr@{}}
\toprule
Method & BLEU & COMET & FRD
& \laalcu{} & Erase \\
\midrule
History-SFT & 30.77 & 85.61 & 2.45 & 3.42 / 6.38 & 1.112 \\
\midrule
Current-draft RL & 31.32 & \textbf{85.72} & 5.07 & 3.40 / 6.48 & 4.514 \\
Hibiki-Zero-style & 30.05 & \underline{85.63} & 2.43 & 3.56 / 6.65 & 1.110 \\
HPO-style & \underline{31.34} & 85.50 & \textbf{2.41} & \underline{3.26} / \underline{6.01} & \underline{1.020} \\
\textbf{PDO} & \textbf{31.58} & 85.37 & \underline{2.42} & \textbf{3.05} / \textbf{5.66} & \textbf{.936} \\
\bottomrule
\end{tabular*}
\end{table}

%% file: tables/cross_domain_results.tex
\begin{table*}[!t]
\centering
\caption{Zero-shot cross-domain evaluation of the FLEURS-trained checkpoints
without target-domain adaptation. Macro reports BLEU, COMET, first-token
latency (FTL), final-response delay (FRD), finalization-aware
computation-unaware length-adaptive average lagging (\laalcu{}; mean/P90), and
normalized erasure. Bold/underline mark best/second-best streaming results;
\laalcu{} mean/P90 are ranked separately. Erasure is shown only for
revision-capable systems.}
\label{tab:crossdomain}
\normalsize
\setlength{\tabcolsep}{0pt}
\renewcommand{\arraystretch}{.82}
\begin{tabular*}{\textwidth}{@{\extracolsep{\fill}}lrrrrrrrrrrcr@{}}
\toprule
\multicolumn{13}{c}{\textbf{(a) Europarl-ST v1.1}} \\
\addlinespace[1pt]
\multirow{2}{*}{\textbf{System}}
& \multicolumn{2}{c}{\textbf{En$\rightarrow$De}}
& \multicolumn{2}{c}{\textbf{En$\rightarrow$Es}}
& \multicolumn{2}{c}{\textbf{En$\rightarrow$Fr}}
& \multicolumn{6}{c}{\textbf{Macro}} \\
\cmidrule(lr){2-3}\cmidrule(lr){4-5}\cmidrule(lr){6-7}\cmidrule(l){8-13}
&
BLEU & COMET & BLEU & COMET & BLEU & COMET
& BLEU & COMET & FTL & FRD & \laalcu & Erase \\
\midrule
\hspace{.7em}SeamlessStreaming
& 21.33 & 76.89 & 32.22 & 81.36 & 24.12 & 78.24
& 25.89 & 78.83 & \underline{2.33} & \textbf{2.47} & \textbf{1.83/2.68} & -- \\
\addlinespace[1pt]
\hspace{.7em}History-SFT
& \underline{26.08} & \underline{84.32}
& \underline{35.23} & \textbf{86.06}
& \underline{26.38} & \textbf{83.18}
& \underline{29.23} & \textbf{84.52} & \textbf{1.98} & 2.54 & 2.77/5.07 & \underline{.735} \\
\addlinespace[1pt]
\hspace{.7em}\textbf{PDO}
& \textbf{27.37} & \textbf{84.44}
& \textbf{36.64} & \underline{85.82}
& \textbf{27.07} & \underline{83.04}
& \textbf{30.36} & \underline{84.44} & \textbf{1.98} & \underline{2.52} & \underline{2.52/4.59} & \textbf{.595} \\
\midrule
\multicolumn{13}{c}{\textbf{(b) CoVoST 2}} \\
\addlinespace[1pt]
\multirow{2}{*}{\textbf{System}}
& \multicolumn{2}{c}{\textbf{En$\rightarrow$Zh}}
& \multicolumn{2}{c}{\textbf{En$\rightarrow$De}}
& \multicolumn{2}{c}{\textbf{En$\rightarrow$Ja}}
& \multicolumn{6}{c}{\textbf{Macro}} \\
\cmidrule(lr){2-3}\cmidrule(lr){4-5}\cmidrule(lr){6-7}\cmidrule(l){8-13}
&
BLEU & COMET & BLEU & COMET & BLEU & COMET
& BLEU & COMET & FTL & FRD & \laalcu & Erase \\
\midrule
\hspace{.7em}SeamlessStreaming
& 31.46 & 79.04 & \textbf{30.56} & 81.63 & 21.49 & 83.03
& 27.84 & 81.23 & \underline{2.88} & 3.07 & \textbf{2.19/3.51} & -- \\
\addlinespace[1pt]
\hspace{.7em}History-SFT
& \underline{39.30} & \underline{84.47}
& 28.39 & \underline{82.04}
& \underline{29.84} & \underline{86.31}
& \underline{32.51} & \underline{84.28} & \textbf{2.00} & \textbf{2.41} & 2.56/4.56 & \underline{.824} \\
\addlinespace[1pt]
\hspace{.7em}\textbf{PDO}
& \textbf{41.03} & \textbf{84.51}
& \underline{30.17} & \textbf{82.08}
& \textbf{30.34} & \textbf{86.36}
& \textbf{33.85} & \textbf{84.32} & \textbf{2.00} & \textbf{2.41} & \underline{2.41/4.21} & \textbf{.753} \\
\bottomrule
\end{tabular*}
\end{table*}

%% file: references.bib
@IEEEtranBSTCTL{IEEEexample:BSTcontrol,
  CTLuse_forced_etal = "yes",
  CTLmax_names_forced_etal = "3",
  CTLnames_show_etal = "3",
  CTLdash_repeated_names = "no",
  CTLuse_doi = "no"
}

@inproceedings{arivazhagan2020retranslation,
  author = {Naveen Arivazhagan and others},
  title = {Re-translation versus Streaming for Simultaneous Translation},
  booktitle = {Proc. {IWSLT}},
  pages = {220--227},
  doi = {10.18653/v1/2020.iwslt-1.27},
  year = {2020}
}

@inproceedings{arivazhagan2020strategies,
  author = {Naveen Arivazhagan and others},
  title = {Re-Translation Strategies for Long Form, Simultaneous, Spoken Language Translation},
  booktitle = {Proc. {IEEE} {ICASSP}},
  pages = {7919--7923},
  doi = {10.1109/ICASSP40776.2020.9054585},
  year = {2020}
}

@inproceedings{ren2020simulspeech,
  author = {Yi Ren and Jinglin Liu and Xu Tan and Chen Zhang and Tao Qin and Zhou Zhao and Tie-Yan Liu},
  title = {{SimulSpeech}: End-to-End Simultaneous Speech to Text Translation},
  booktitle = {Proc. {ACL}},
  pages = {3787--3796},
  doi = {10.18653/v1/2020.acl-main.350},
  year = {2020}
}

@inproceedings{ma2020simulmt,
  author = {Xutai Ma and Juan Pino and Philipp Koehn},
  title = {{SimulMT} to {SimulST}: Adapting Simultaneous Text Translation to End-to-End Simultaneous Speech Translation},
  booktitle = {Proc. {AACL}},
  pages = {582--587},
  doi = {10.18653/v1/2020.aacl-main.58},
  year = {2020}
}

@inproceedings{papi2023attention,
  author = {Sara Papi and Matteo Negri and Marco Turchi},
  title = {Attention as a Guide for Simultaneous Speech Translation},
  booktitle = {Proc. {ACL}},
  pages = {13340--13356},
  doi = {10.18653/v1/2023.acl-long.745},
  year = {2023}
}

@inproceedings{papi2023alignatt,
  author = {Sara Papi and others},
  title = {{AlignAtt}: Using Attention-based Audio-Translation Alignments as a Guide for Simultaneous Speech Translation},
  booktitle = {Proc. Interspeech},
  pages = {3974--3978},
  doi = {10.21437/Interspeech.2023-170},
  year = {2023}
}

@inproceedings{papi2024simulseamless,
  author = {Sara Papi and others},
  title = {{SimulSeamless}: {FBK} at {IWSLT} 2024 Simultaneous Speech Translation},
  booktitle = {Proc. {IWSLT}},
  pages = {120--127},
  doi = {10.18653/v1/2024.iwslt-1.11},
  year = {2024}
}

@inproceedings{labiausse2026hibikizero,
  author = {Tom Labiausse and Romain Fabre and Yannick Est{\`e}ve and Alexandre D{\'e}fossez and Neil Zeghidour},
  title = {Simultaneous Speech-to-Speech Translation Without Aligned Data},
  booktitle = {Proc. {ICML}},
  series = {Proc. Mach. Learn. Res.},
  volume = {306},
  year = {2026}
}

@inproceedings{ouyang2026hpo,
  author = {Siqi Ouyang and Shuoyang Ding and Oleksii Hrinchuk and Vitaly Lavrukhin and Brian Yan and Boris Ginsburg and Lei Li},
  title = {Hierarchical Policy Optimization for Simultaneous Translation of Unbounded Speech},
  booktitle = {Proc. {ACL}},
  pages = {1772--1787},
  doi = {10.18653/v1/2026.acl-long.80},
  year = {2026}
}

@inproceedings{sen2023selftraining,
  author = {Sukanta Sen and others},
  title = {Self-training Reduces Flicker in Retranslation-based Simultaneous Translation},
  booktitle = {Proc. {EACL}},
  pages = {3734--3744},
  doi = {10.18653/v1/2023.eacl-main.270},
  year = {2023}
}

@inproceedings{papi2022laal,
  author = {Sara Papi and others},
  title = {Over-Generation Cannot Be Rewarded: Length-Adaptive Average Lagging for Simultaneous Speech Translation},
  booktitle = {Proc. {AutoSimTrans}},
  pages = {12--17},
  doi = {10.18653/v1/2022.autosimtrans-1.2},
  year = {2022}
}

@article{shi2026qwen3asr,
  author = {Xian Shi and Xiong Wang and Zhifang Guo and Yongqi Wang and Pei Zhang and Xinyu Zhang and Zishan Guo and Hongkun Hao and Yu Xi and Baosong Yang and Jin Xu and Jingren Zhou and Junyang Lin},
  title = {{Qwen3-ASR} Technical Report},
  journal = {arXiv:2601.21337},
  year = {2026}
}

@inproceedings{hu2022lora,
  author = {Edward J. Hu and Yelong Shen and Phillip Wallis and Zeyuan Allen-Zhu and Yuanzhi Li and Shean Wang and Lu Wang and Weizhu Chen},
  title = {{LoRA}: Low-Rank Adaptation of Large Language Models},
  booktitle = {Proc. {ICLR}},
  year = {2022}
}

@inproceedings{lin2004rouge,
  author = {Chin-Yew Lin},
  title = {{ROUGE}: A Package for Automatic Evaluation of Summaries},
  booktitle = {Text Summarization Branches Out},
  pages = {74--81},
  year = {2004}
}

@inproceedings{post2018sacrebleu,
  author = {Matt Post},
  title = {A Call for Clarity in Reporting {BLEU} Scores},
  booktitle = {Proc. {WMT}},
  pages = {186--191},
  doi = {10.18653/v1/W18-6319},
  year = {2018}
}

@article{schulman2017ppo,
  author = {John Schulman and others},
  title = {Proximal Policy Optimization Algorithms},
  journal = {arXiv:1707.06347},
  year = {2017}
}

@article{shao2024deepseekmath,
  author = {Zhihong Shao and others},
  title = {{DeepSeekMath}: Pushing the Limits of Mathematical Reasoning in Open Language Models},
  journal = {arXiv:2402.03300},
  year = {2024}
}

@article{seamless2023,
  author = {{Seamless Communication} and others},
  title = {{Seamless}: Multilingual Expressive and Streaming Speech Translation},
  journal = {arXiv:2312.05187},
  year = {2023}
}

@inproceedings{machacek2025simulstreaming,
  author = {Dominik Mach{\'a}{\v{c}}ek and Peter Pol{\'a}k},
  title = {Simultaneous Translation with Offline Speech and {LLM} Models in {CUNI} Submission to {IWSLT} 2025},
  booktitle = {Proc. {IWSLT}},
  pages = {389--398},
  doi = {10.18653/v1/2025.iwslt-1.41},
  year = {2025}
}

@article{fu2026easist,
  author = {Biao Fu and Donglei Yu and Minpeng Liao and Chengxi Li and Xinjie Chen and Yidong Chen and Kai Fan and Xiaodong Shi},
  title = {Efficient and Adaptive Simultaneous Speech Translation with Fully Unidirectional Architecture},
  journal = {Proc. {AAAI}},
  volume = {40},
  number = {36},
  pages = {30735--30743},
  doi = {10.1609/aaai.v40i36.40330},
  year = {2026}
}

@article{koshkin2026hikari,
  author = {Roman Koshkin and others},
  title = {Streaming Translation and Transcription Through Speech-to-Text Causal Alignment},
  journal = {arXiv:2603.11578},
  year = {2026}
}

@inproceedings{fuxa2026alignatt4llm,
  author = {Quentin Fuxa and Dominik Mach{\'a}{\v{c}}ek},
  title = {{AlignAtt4LLM}: Fast {AlignAtt} for Decoder-Only {LLM}s at {IWSLT} 2026 Simultaneous Speech Translation Task},
  booktitle = {Proc. {IWSLT}},
  pages = {284--295},
  doi = {10.18653/v1/2026.iwslt-1.32},
  year = {2026}
}

@article{he2026simuls2stomni,
  author = {Rongshen He and Xinyu Liang and Dekun Chen and Jiaqi Li and Mingjie Chen and Zhizheng Wu},
  title = {{SimulS2ST-Omni}: Data-Efficient Streaming Speech-to-Speech Translation via Explicit Trajectory Supervision},
  journal = {arXiv:2607.19810},
  year = {2026}
}

@article{zheng2026hymt2,
  author = {Mao Zheng and Zheng Li and Tao Chen and Bo Lv and Mingrui Sun and Mingyang Song and Jinlong Song and Hong Huang and Decheng Wu and Hai Wang and Yifan Song and Yanfeng Chen and Guanwei Zhang},
  title = {{Hy-MT2}: A Family of Fast, Efficient and Powerful Multilingual Translation Models in the Wild},
  journal = {arXiv:2605.22064},
  year = {2026}
}

@inproceedings{conneau2022fleurs,
  author = {Alexis Conneau and others},
  title = {{FLEURS}: Few-shot Learning Evaluation of Universal Representations of Speech},
  booktitle = {Proc. {IEEE} {SLT}},
  pages = {798--805},
  doi = {10.1109/SLT54892.2023.10023141},
  year = {2023}
}

@inproceedings{iranzo2020europarlst,
  author = {Javier Iranzo-S{\'a}nchez and others},
  title = {{Europarl-ST}: A Multilingual Corpus for Speech Translation of Parliamentary Debates},
  booktitle = {Proc. {IEEE} {ICASSP}},
  pages = {8229--8233},
  doi = {10.1109/ICASSP40776.2020.9054626},
  year = {2020}
}

@inproceedings{wang2020covost2,
  author = {Changhan Wang and Anne Wu and Jiatao Gu and Juan Pino},
  title = {{CoVoST} 2 and Massively Multilingual Speech Translation},
  booktitle = {Proc. Interspeech},
  pages = {2247--2251},
  doi = {10.21437/Interspeech.2021-2027},
  year = {2021}
}

@inproceedings{popovic2017chrf,
  author = {Maja Popovi{\'c}},
  title = {{chrF++}: Words Helping Character n-grams},
  booktitle = {Proc. {WMT}},
  pages = {612--618},
  doi = {10.18653/v1/W17-4770},
  year = {2017}
}

@inproceedings{rei2022comet22,
  author = {Ricardo Rei and others},
  title = {{COMET}-22: Unbabel-{IST} 2022 Submission for the Metrics Shared Task},
  booktitle = {Proc. {WMT}},
  pages = {578--585},
  doi = {10.18653/v1/2022.wmt-1.52},
  year = {2022}
}

@inproceedings{radford2022whisper,
  author = {Alec Radford and Jong Wook Kim and Tao Xu and Greg Brockman and Christine Mcleavey and Ilya Sutskever},
  title = {Robust Speech Recognition via Large-Scale Weak Supervision},
  booktitle = {Proc. {ICML}},
  series = {Proc. Mach. Learn. Res.},
  volume = {202},
  pages = {28492--28518},
  year = {2023}
}

@article{martins2025eurollm,
  author = {Pedro Henrique Martins and others},
  title = {{EuroLLM}: Multilingual Language Models for {Europe}},
  journal = {Procedia Comput. Sci.},
  volume = {255},
  pages = {53--62},
  doi = {10.1016/j.procs.2025.02.260},
  year = {2025}
}
